\documentclass[final,numberedheadings]{aipproc}%
\usepackage{amsmath}
\usepackage{amsfonts}
\usepackage{amssymb}
\usepackage{graphicx}
 \layoutstyle{8x11single}
\begin{document}


\title{Special Relativity and Reactions with Unstable Nuclei}
\author{C.A. Bertulani}
{ address={Department of Physics, University of Arizona, Tucson, Arizona
85721} }

\begin{abstract}
Dynamical relativistic effects are often neglected in the
description of reactions with unstable nuclear beams at intermediate
energies ($E_{Lab}\approx100$ MeV/nucleon). Evidently, this
introduces sizable errors in experimental analysis and theoretical
descriptions of these reactions. This is particularly important for
the experiments held in GANIL/France, MSU/USA, RIKEN/Japan and
GSI/Germany. I review a few examples where relativistic effects have
been studied in nucleus-nucleus scattering at intermediate energies.
\end{abstract}
\maketitle

\section{Introduction}

The number of radioactive beam facilities are growing fast around
the world. Some of these facilities use the fragmentation technique,
with secondary beams in the energy range $E_{Lab}\approx100$
MeV/nucleon. Examples are the facilities in GANIL/France, MSU/USA,
RIKEN/Japan and GSI/Germany. Relativity, an obviously important
physics concept \cite{Ei1905}, is often neglected in calculations
aiming at relating the reaction mechanisms to the internal structure
of the projectiles. For example, popular DWBA codes (FRESCO, ECIS,
DWUCK, etc.) useful in the analysis of nuclear reactions, include
relativity only in kinematic relations. The effects of relativity in
the reaction dynamics (i.e. the interaction) is not accounted for
because these codes were intended for lower energies. It is also
important to notice that the inclusion of relativistic effects in
the nucleus-nucleus dynamics is a very difficult task. A fully
covariant treatment of the nuclear many-body scattering (with
inclusion of retardation effects between all nucleons) is not
possible without approximations.

In this short article I will review a few examples where
relativistic effects  have been included in nuclear reactions at
intermediate energies. It is worthwhile to observe that after 100
years of relativity \cite{Ei1905} this still remains a challenge in
many aspects.

\section{Semiclassical methods and elastic scattering}

Semiclassical methods are a very popular tool for the description of
nucleus-nucleus collisions at high energies. As an example, I cite
the Coulomb excitation mechanism, in which the inelastic cross
section can be factorized as
\begin{equation}
{\frac{d\sigma_{i\rightarrow f}}{d\Omega}}=\left(
\frac{d\sigma}{d\Omega }\right)  _{\mathrm{el}}P_{i\rightarrow f},
\end{equation}
where $P_{i\rightarrow f}$\ is the probability for a nuclear
transition between the states $i$ and $f$ when the nuclei scatter
through an angle $\Omega$. The collision dynamics enters here in two
distinct ways: in the calculation of $P_{i\rightarrow f}$ and
$\left(  d\sigma/d\Omega\right)  _{\mathrm{el}}$. Let us forget
about $P_{i\rightarrow f}$ for a moment and let us investigate
$\left( d\sigma/d\Omega\right) _{\mathrm{el}}$. The question here is
what is the error done by  measuring the number of particles
scattered to $\Omega$ and using the Rutherford formula for $\left(
d\sigma /d\Omega\right)  _{\mathrm{el}}$. This has been investigated
in ref. \cite{AAB90}.

A system of two point charges interacting electromagnetically and
moving at low velocities can be described by an approximate
Lagrangian which depends only on the degrees of freedom of the
particles neglecting those related to the electromagnetic field (the
Darwin Lagrangian's). In this approximation it is possible to
separate the degrees of freedom associated with the relative
position $r$ and relative velocity $v$ of the particles from the
center of mass degrees of freedom. For a system of particles with
different masses this approximation is only possible up to  the
$c^{-2}$ order, whereas for a system with equal charge- to-mass
ratio (with $Z_{1}e/m_{1}=Z_{2}e/m_{2}$ ) the approximation goes up
to order $c^{-4}$. Using Lagrange's equation of motion, it is then
straightforward to obtain a numerical result for the deflection
angle and the elastic cross section. Explicit expressions for these
Lagrangians are given in ref. \cite{AAB90}. The authors show that
the scattering angle increases by up to 6\% when relativistic
corrections are included in $^{208}\mathrm{Pb}+^{208}\mathrm{Pb}$
collisions at 100 MeV/nucleon. The effect on the elastic scattering
cross section is even more drastic: $\approx13\%$ for center-of-mass
scattering angles around 0-4 degrees.

Another result obtained in ref. \cite{AAB90} is that the effects of
relativity can be taken into account in a much simpler way when the
projectile
is a light particle scattering on a heavy target (e.g., $^{11}\mathrm{Li}%
+^{208}\mathrm{Pb}$). In this case, the main contribution of
relativistic corrections (up to 80\% of the total effect) is due to
the changes in masses, an easily implementable correction. Only
about 20\% of the relativistic effects are due to magnetic
interactions and retardation. Then, a good approximation for the
elastic scattering is given by
\begin{equation}
\left(  \frac{d\sigma}{d\Omega}\right)  _{\mathrm{el}}\left(  \beta
,\Theta\right)  =\left[  \frac{Z_{1}Z_{2}e^{2}}{2mc^{2}\beta^{2}\sin
^{2}\left(  \Theta/2\right)  }\right]  ^{2}\left[  1-h\left(  \Theta\right)
\beta^{2}+\mathcal{O}\left(  \beta^{2}\right)  \right]
\end{equation}
where%
\[
h\left(  \Theta\right)  =1+\frac{1}{2}\left[  1+\left(  \pi-\Theta\right)
\cot\Theta\right]  \tan^{2}\frac{\Theta}{2},\ \ \ \ \ \ \ \ \ \Theta
=\pi-2\frac{\eta}{\sqrt{\eta^{2}-\beta^{2}}}\arctan\sqrt{\eta^{2}-\beta^{2}},
\]
with $\eta=\beta Lc/Z_{1}Z_{2}e^{2},$ $L$ being the angular momentum
of the system, and $\beta=v/c$. The function $h\left(  \Theta\right)
$ is always positive for the relevant scattering angles. Therefore
one concludes that the relativistic corrections in the elastic
Coulomb cross section are always negative if they are parametrized
as a function of the initial velocity. On the other hand if one
fixes the initial kinetic energy, instead of the velocity, the
corrections will be always positive, as shown in ref. \cite{AAB90}.

\begin{figure}[ptb]
\includegraphics[height=.35\textheight]{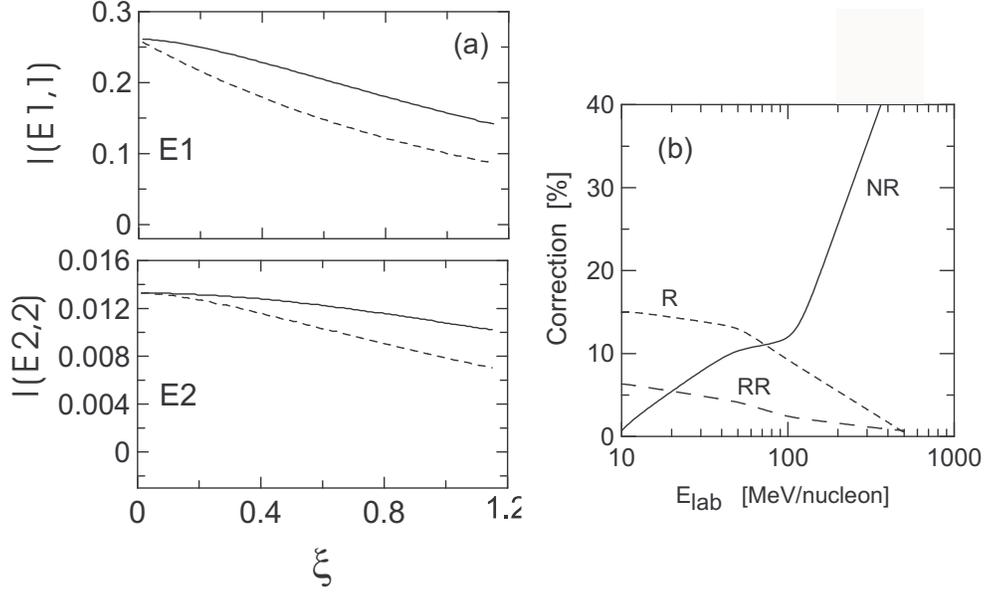}
\caption{(a) Upper panel: Real  part (solid line) of the orbital integral
$I(E1,1)$ for $\gamma=1.1$ ($E_{\mathrm{lab}%
}\simeq100$ MeV/nucleon), calculated with a theory containing both
retardation and relativistic corrections to the Coulomb trajectory.
The approximation used in ref. \cite{WA79}, neglecting relativistic
corrections to the Coulomb trajectory, is shown by the dashed line.
Lower panel: Same plot, but for the orbital integral $I(E2,2)$. (b)
Percentage correction in the calculation of the cross sections for
Coulomb excitation of the 0.89 MeV state in $^{40}$S + $^{197}$Au
collisions as a function of the bombarding energy. The solid line
(NR) corresponds to the use of the non-relativistic orbital
integrals \cite{AW75} compared to the "exact" calculation of ref.
\cite{Ber03}. The same is plotted for the other two cases: (R) with
retardation effects only \cite{WA79}, and (RR) with the retardation
effects plus an approximate recoil
correction \cite{WA79}. For more details, see ref. \cite{WA79}. }%
\label{f1}%
\end{figure}

The study mentioned above is useful for an analysis of experimental
data on inelastic nucleus-nucleus scattering in intermediate energy
collisions. The extension of these calculations to all orders in
$\beta$ is an intriguing problem and an appropriate tool could be
the use of the action-at-a-distance electrodynamics of Fokker,
Wheeler and Feynman \cite{FWF45}. Certainly, more studies in this
direction are needed and the present focus on nuclear reactions at
intermediate energies seem to be a good opportunity to investigate
such effects.

\section{Another tractable case: Coulomb excitation}

The theory of Coulomb excitation in low-energy collisions is very
well understood~\cite{AW75}. It has been used and improved for over
thirty years to infer electromagnetic properties of nuclei and has
also been tested in experiments to a high level of precision. A
large number of small corrections are now well known in the theory
and are necessary in order to analyze experiments on multiple
excitation and reorientation effects.

In the case of relativistic heavy ion collisions pure Coulomb excitation may
be distinguished from the nuclear reactions by demanding extreme forward
scattering or avoiding the collisions in which violent reactions take place.
The Coulomb excitation of relativistic heavy ions is thus characterized by
straight-line trajectories with impact parameter $b$ larger than the sum of
the radii of the two colliding nuclei. A detailed calculation of relativistic
electromagnetic excitation on this basis was performed by Winther and
Alder~\cite{WA79}. As in the non-relativistic case, they showed how one can
separate the contributions of the several electric ($E\lambda$) and magnetic
($M\lambda$) multipolarities to the excitation. Later, it was shown that a
quantum theory for relativistic Coulomb excitation leads to minor
modifications of the semiclassical results~\cite{BB88}. In
Refs.~\cite{AB89,BN93} the inclusion of relativistic effects in semiclassical
and quantum formulations of Coulomb excitation was fully clarified.

Recently, the importance of relativistic effects in Coulomb
excitation of a projectile by a target with charge $Z_{2}$, followed
by gamma-decay in nuclear reactions at intermediate energies was
studied in details. The Coulomb excitation cross section is given by
\begin{equation}
{\frac{d\sigma_{i\rightarrow f}}{d\Omega}}=\left(
\frac{d\sigma}{d\Omega }\right)
_{\mathrm{el}}\frac{16\pi^{2}Z_{2}^{2}e^{2}}{\hbar^{2}}\sum
_{\pi\lambda\mu}{\dfrac{B(\pi\lambda,I_{i}\rightarrow
I_{f})}{(2\lambda
+1)^{3}}}\mid S(\pi\lambda,\mu)\mid^{2},\label{cross_2}%
\end{equation}
where $B(\pi\lambda,I_{i}\rightarrow I_{f})$ is the reduced
transition probability of the projectile nucleus, $\pi\lambda=E1,\
E2,\ M1,\ldots$ is the multipolarity of the excitation, and
$\mu=-\lambda,-\lambda+1,\ldots,\lambda$. The orbital integrals
$S(\pi\lambda,\mu$) contain the information about relativistic
corrections on the relative motion between the nuclei, as well as
the relativistic effects on the excitation mechanism (e.g. retarded
Coulomb interaction). Inclusion of absorption effects in
$S(\pi\lambda,\mu$) due to the imaginary part of an optical
nucleus-nucleus potential where worked out in ref. \cite{BN93}.
These orbital integrals depend on the Lorentz factor
$\gamma=(1-v^{2}/c^{2})^{-1/2}$, with $c$ being the speed of light,
on the multipolarity $\pi\lambda\mu$, and on the adiabacity
parameter $\xi (b)=\omega_{fi}b/\gamma v<1$, where
$\omega_{fi}=\left( E_{f}-E_{i}\right) /\hbar$ is the excitation
energy (in units of $\hbar$) and $b$ is the impact parameter. It is
often more convenient to write the orbital integrals in another
form, e.g. $S(E\lambda,\mu)
=\dfrac{\mathcal{C}_{\lambda\mu}}{va^{\lambda}}\; I(E\lambda,\mu)$,
where $\mathcal{C}_{\lambda\mu}$ is a geometrical factor, depending
only on $\lambda$ and $\mu$, $a=Z_{1}Z_{2}e^{2}/\gamma mv^{2}$, and
$I(E\lambda,\mu)$ is now the orbital integral in an dimensionless
form.

Figure 1(a) (from \cite{Ber03}) shows the effect of relativity on
the Coulomb interaction  and classical trajectory corrections in
nuclear collisions at intermediate collisions. The comparison is
made in terms of the variable $\xi$\, which is the appropriate
variable for high energy collisions. Only for $\xi\ll1$ the
expressions in the relativistic limit reproduce the correct behavior
of the orbital integrals.

\begin{table}[th]
\begin{tabular}
[c]{|c|c|c|c|c|c|c|}\hline
Nucleus & E$_{x}$ & B(E2) & 10 MeV/A & 50 MeV/A & 100 MeV/A & 500 MeV/A\\
& [MeV] & [e$^{2}$fm$^{4}$] & $\sigma_{C}$ [mb] & $\sigma_{C}$ [mb] &
$\sigma_{C}$ [mb] & $\sigma_{C}$ [mb]\\\hline
$^{38}$S & 1.29 & 235 & (492) 500 [651] & (80.9) 91.7 [117] & (40.5) 50.1
[57.1] & (9.8) 16.2 [16.3]\\\hline
$^{40}$S & 0.89 & 334 & (877) 883 [1015] & (145.3) 162 [183] & (76.1) 85.5
[93.4] & (9.5) 20.9 [21.]\\\hline
$^{42}$S & 0.89 & 397 & (903) 908 [1235] & (142.7) 158 [175] & (65.1) 80.1
[89.4] & (9.9) 23.2 [23.4]\\\hline
$^{44}$Ar & 1.14 & 345 & (747) 752 [985] & (133) 141 [164] & (63.3) 71.7
[80.5] & (8.6) 17.5 [17.6]\\\hline
$^{46}$Ar & 1.55 & 196 & (404) 408 [521] & (65.8) 74.4 [88.5] & (30.2) 37.4
[41.7] & (5.72)\ 10.8 [11]\\\hline
\end{tabular}
\caption{ Coulomb excitation cross sections of the first excited state in
$^{38,40,42}$S and $^{44,46}$Ar projectiles at 10, 50 100 and 500 MeV/nucleon
incident on gold targets. The numbers inside parenthesis and brackets were
obtained with pure non-relativistic and straight-line relativistic
calculations, respectively. The numbers at the center are obtained with the
"full" account of relativistic effects, as explained in ref. \cite{Ber03}.}%
\label{tab:Table_1}%
\end{table}

Table~\ref{tab:Table_1} shows the effects of relativistic
corrections in the collision of the radioactive nuclei
$^{38,40,42}$S and $^{44,46}$Ar on gold targets. These reactions
have been studied at $E_{lab}\sim40$ MeV/nucleon at the MSU
facility~\cite{Sch96}.  Table~\ref{tab:Table_1} shows the Coulomb
excitation cross sections of the first excited state in each nucleus
as a function of the bombarding energy per nucleon. The cross
sections are given in milibarns. The numbers inside parenthesis and
brackets were obtained with pure non-relativistic \cite{AW75} and
relativistic calculations \cite{WA79}, respectively. The minimum
impact parameter is chosen so that the distance of closest approach
corresponds to the sum of the nuclear radii in a collision following
a Rutherford trajectory. One observes that at 10 MeV/nucleon the
relativistic corrections are important only at the level of 1\%. At
500 MeV/nucleon, the correct treatment of the recoil corrections  is
relevant on the level of 1\%. Thus the non-relativistic treatment of
Coulomb excitation~\cite{AW75} can be safely used for energies below
about 10 MeV/nucleon and the relativistic treatment with a
straight-line trajectory~\cite{WA79} is adequate above about 500
MeV/nucleon. However at energies around 50 to 100 MeV/nucleon,
accelerator energies common to most radioactive beam facilities
(MSU, RIKEN, GSI, GANIL), it is very important to use a correct
treatment of recoil and relativistic effects, both kinematically and
dynamically. At these energies, the corrections can add up to 50\%.
These effects were also shown in Ref.~\cite{AB89} for the case of
excitation of giant resonances in collisions at intermediate
energies.

We conclude that a reliable extraction useful nuclear properties,
like the electromagnetic response (B(E2)-values, $\gamma$-ray
angular distribution, etc.) from Coulomb excitation experiments at
intermediate energies requires a proper treatment of special
relativity. The effect is highly non-linear, i.e. a 10\% increase in
the velocity might lead to a 50\% increase (or decrease) of certain
physical quantities.

\section{Strong interaction: Glauber, DWBA and semiclassical methods}

The treatment of the strong interaction in nucleus-nucleus
collisions at intermediate and high energies is evidently much more
complicated than the case of Coulomb excitation, described in the
previous section. Fortunately, many direct nuclear processes, e.g.
nucleon knockout, or stripping, elastic breakup (diffraction
dissociation), etc, are possible to study using the optical limit of
the Glauber theory, in which the nuclear ground-state densities and
the nucleon-nucleon total cross sections are the main input. In
fact, this method has become one of the main tools in the study of
nuclei far from stability \cite{han03}. The reason is that the
eikonal (or Glauber) methods only use the dependence of the
scattering matrices, $S(b)$, on the transverse direction, $b$.
Transverse directions are always Lorentz invariants. The reason for
using $S(b)$, instead of $S({\bf r})$, traces back to the eikonal
scattering wavefunction at the asymptotic region ($r\longrightarrow
\infty$),
\begin{equation}
\Psi_{scatt}= S\left( b\right) \ \exp\left( i\mathbf{k}\cdot
\mathbf{r}\right) \ , \label{eik}
\end{equation}
where ${\bf k}$ is the particle's momentum, and ${\bf r}$ its
position. Obviously, the plane wave part of eq. \ref{eik}  is
Lorentz invariant. The S-matrix in eq. \ref{eik} is
$S=\exp\{i\chi(b)\}$, where $\chi(b)$ is the eikonal phase-shift,
given in terms of the interaction potential $V$ by
\begin{equation}
\chi=-\frac{1}{\hbar v}\int_{-\infty}^{\infty}dz\ V\left( r\right) \ .
\label{eikphase}
\end{equation}
Under a Lorentz transformation to the target system, the coordinate $z$ transforms
as $z\longrightarrow \gamma z$. Thus, strictly speaking, the S-matrix $S(b)$ is
Lorentz invariant only if $V$ transforms as the time-component of a four-vector, i.e.
$V(r) \longrightarrow \gamma V(b,\gamma z)$.

The relativistic property described above is most easily seen within a folding
potential model for a nucleon-nucleus collision:
\begin{equation}
V\left( \mathbf{r}\right) =\int dr^{\prime3}\ \rho_{T}\left( \mathbf{r}%
^{\prime}\right) \ v_{NN}\left( \mathbf{r-r}^{\prime}\right) ,   \label{fold}
\end{equation}
where $\rho_{T}\left( \mathbf{r}^{\prime}\right) $ is the nuclear density of
the target. In the frame of reference of the projectile, the density of the
target looks contracted and particle number conservation leads to the
relativistic modification of eq. \ref{fold} so that $\rho_{T}\left( \mathbf{r%
}^{\prime}\right) \rightarrow\gamma\rho_{T}\left( \mathbf{r}_{\perp}^{\prime}%
\mathbf{,\gamma}z^{\prime}\right) $, where
$\mathbf{r}_{\perp}^{\prime}$ is the transverse component of
\textbf{r'}. But the number of nucleons as seen by the target (or
projectile) per unit area remains the
same. In other words, a change of variables $z^{\prime\prime }=\mathbf{\gamma%
}z^{\prime}$\ \ in the integral of eq. \ref{fold} seems to restore
the same eq. \ref{fold}. However, this change of variables also
modifies the nucleon-nucleon interaction $v_{NN}$. Thus, relativity
introduces non-trivial effects in a potential model description of
nucleus-nucleus scattering at high energies. Colloquially speaking,
nucleus-nucleus scattering at high energies is not simply an
incoherent sequence of nucleon-nucleon collisions. Since the
nucleons are confined within a box (inside the nucleus), Lorentz
contraction induces a collective effect: in the extreme limit
$\gamma\rightarrow\infty$ all nucleons would interact at once with
the projectile. This is often neglected in pure geometrical (Glauber
model) description of nucleus-nucleus collisions at high energies,
as it is assumed that the nucleons inside \textquotedblleft
firetubes" scatter independently.

Assuming that the nucleon-nucleon interaction is of very short range so that
the approximation $\ v_{NN}\left( \mathbf{r-r}^{\prime}\right) =J_{0}\
\delta\left( \mathbf{r-r}^{\prime}\right) $\ can be used, one sees from eq. %
\ref{fold} that $V\left( \mathbf{r}\right) $, the interaction that a
nucleon in the projectile has with the target nucleus, also has
similar transformation properties as the density: $V\left(
\mathbf{r}\right) \rightarrow\mathbf{\gamma}V\left(
\mathbf{r}_{\perp}\mathbf{,\gamma}z\right) $, i.e. $V\left(
\mathbf{r}\right) $ transforms as the time-component of a
four-vector. In this situation, the Lorentz contraction has no
effect whatsoever in the diffraction dissociation amplitudes,
described in the previous sections within the eikonal approximation.
This is because a change of variables $z^{\prime}=\mathbf{\gamma}z$\
in the eikonal phases leads to the same result as in the
non-relativistic case, as can be easily checked from eq.
\ref{eikphase}. Of course, the delta-function approximation for the
nucleon-nucleon interaction means that nucleons will scatter at
once, and Lorentz contraction does not introduce any additional
collective effect. This is not the case for realistic interactions
with finite range and collisions at intermediate energies.

Using the eikonal approach to account for scattering of particles
within the projectile, one obtains the wavefunctions
for the initial and final states as%
\begin{equation}
\Psi_{i}=\phi_{i}\left( \mathbf{r}\right) \exp\left( i\mathbf{k}\cdot\mathbf{%
R}\right) ,\ \ \ \ \ \ \ \ \ \ \ \ \Psi_{f}=\phi_{f}\left( \mathbf{r}\right)
S\left( b\right) \ \exp\left( i\mathbf{k}\cdot \mathbf{R}\right) \ ,
\end{equation}
where $\phi_{i,f}\left( \mathbf{r}\right) $ are the initial and final
probability amplitudes (wavefunctions) that a particle in the projectile is
at a distance $\mathbf{r}$ from its center of mass. The particle's $S$%
-matrix, $S\left( b\right) $, accounts for the distortion due to the
interaction. If we now assume that the projectile is a two-body
system (e.g. a core+valence particle), we get the {\it diffraction
dissociation} formula as follows. The wavefunction of a two-body
projectile in the
initial and final states is given by%
\begin{equation}
\Psi_{i}  =\phi_{i}\left( \mathbf{r}\right) \exp\left[ i\left( \mathbf{k}%
_{c}\cdot\mathbf{r}_{c}+\mathbf{k}_{v}\cdot\mathbf{r}_{v}\right)
\right], \ \ \ \ \ \ \ \ \ \ \ \Psi_{f}  =\phi_{f}\left(
\mathbf{r}\right) S_{c}\left( b_{c}\right)
S_{v}\left( b_{v}\right) \exp\left[ i\left( \mathbf{k}_{c}^{\prime}\cdot%
\mathbf{r}_{c}+\mathbf{k}_{v}^{\prime}\cdot\mathbf{r}_{v}\right) \right] \ ,
\label{eikw1}
\end{equation}
where now $\phi_{i,f}\left( \mathbf{r}\right) $ are the initial and final
intrinsic wavefunctions of the (core+valence particle) as a function of \ $%
\mathbf{r=r}_{1}-\mathbf{r}_{2}$. The relation between the intrinsic, $%
\mathbf{r}$, and center of mass, $\mathbf{R}$, coordinates is given in terms
of the mass ratios $\beta_{i}=m_{i}/m_{P}$. Explicitly, $\mathbf{r}_{v}=%
\mathbf{R}+\beta_{c}\mathbf{r}$ and $\mathbf{r}_{c}=\mathbf{R}-\beta _{v}%
\mathbf{r}$. The core and valence particle $S$-matrices, $S_{c}\left(
b_{c}\right) $ and$\ S_{v}\left( b_{v}\right) $, account for the distortion
due to the interaction with the target.

The probability amplitude for diffraction dissociation is the overlap
between the two wavefunctions above, i.e.
\begin{equation}
A_{\mathrm{(diff)}}=\int d^{3}r_{c}d^{3}r_{v}\ \phi_{f}^{\ast}\left( \mathbf{%
r}\right) \phi_{i}\left( \mathbf{r}\right) \delta\left( z_{c}+z_{v}\right)
S_{c}\left( b_{c}\right) S_{v}\left( b_{v}\right) \exp\left[ i\left( \mathbf{%
q}_{c}\cdot\mathbf{r}_{c}+\mathbf{q}_{v}\cdot\mathbf{r}_{v}\right) \right] ,
\label{dif1}
\end{equation}
where $\mathbf{q}_{c}=\mathbf{k}_{c}^{\prime}-\mathbf{k}_{c}$ is the
momentum transfer to the core particle, and accordingly for the valence
particle. The above formula yields the probability amplitude that the\
projectile starts the collision as a bound state and ends up as two
separated pieces, in this case, the core and the valence particle (e.g. a
proton or a neutron). All the information for the dissociation mechanism
comes from the knowledge of the $S$-matrices, $S_{c}$ and $S_{v}$. The
delta-function $\delta\left( Z\right) $\ was introduced in the eq. \ref{dif1}
to account for the fact that the $S$-matrices calculated in the eikonal
approximation only depend on the transverse direction.

In the weak interaction limit, or perturbative limit, the
phase-shifts are
very small so that%
\begin{align}
S_{c}\left( b_{c}\right) S_{v}\left( b_{v}\right) & =\exp\left[ i\left(
\chi_{c}+\chi_{v}\right) \right] \ \simeq1+i\chi_{c}+i\chi _{v}  \notag \\
& =1-\frac{i}{\hbar v}\int V_{cT}\left( \mathbf{r}_{c}\right) \ dz_{c}-\frac{%
i}{\hbar v}\int V_{vT}\left( \mathbf{r}_{v}\right) \ dz_{v}.   \label{dwba1}
\end{align}
The factor 1 does not contribute to the breakup\textbf{. }Thus,
inserting
the result above in eq. \ref{dif1}, we obtain%
\begin{equation}
A_{\mathrm{(PWBA)}}\simeq\frac{1}{i\hbar v}\int d^{3}r_{c}d^{3}r_{v}\ \phi
_{f}^{\ast}\left( \mathbf{r}\right) \phi_{i}\left( \mathbf{r}\right) \left[
V_{cT}\left( \mathbf{r}_{c}\right) +V_{vT}\left( \mathbf{r}_{v}\right) %
\right] \exp\left[ i\left( \mathbf{q}_{c}\cdot\mathbf{r}_{c}+\mathbf{q}%
_{v}\cdot\mathbf{r}_{v}\right) \right] ,   \label{dwba2}
\end{equation}
where the integrals over $z_{c}$ and $z_{v}$ in eq. \ref{dwba1} were
absorbed back to the integrals over $\mathbf{r}_{c}$ and $\mathbf{r}_{v}$
after use of the delta-function $\delta\left( z_{c}+z_{v}\right) $. The
above equation is nothing more than the plane-wave Born-approximation (PWBA)
amplitude. However, absorption is not treated properly. For small values of $%
\mathbf{r}_{c}$ and $\mathbf{r}_{v}$ the phase-shifts are not small
and the approximation used in eq. \ref{dwba1} fails.\ A better
approximation is to assume that for small distances, where
absorption is important, $S_{c}\left( b_{c}\right) S_{v}\left(
b_{v}\right) \simeq S\left( b\right) $, where the right-hand side is
the $S$-matrix for the projectile scattering as a whole on the
target. Using the coordinates $\mathbf{r}$ and$\ \mathbf{R}$ , and
defining $U_{int}(\mathbf{r,R})=V_{cT}\left( \mathbf{r}_{c}\right) \
+V_{nT}\left( \mathbf{r}_{n}\right) $, one gets for the T-matrix
\begin{equation}
T_{\mathrm{(DWBA)}}=i\hbar vA_{\mathrm{(DWBA)}}\simeq\int
d^{3}rd^{3}R\
\phi_{f}^{\ast}\left( \mathbf{r}\right) \exp\left[ i\mathbf{q}\cdot\mathbf{r}%
\right] \phi_{i}\left( \mathbf{r}\right) U_{int}(\mathbf{r,R})S\left(
b\right) \exp\left[ i\mathbf{Q}\cdot\mathbf{R}\right] \ .   \label{TDWBA0}
\end{equation}

In elastic scattering, or excitation of collective modes (e.g. giant
resonances), the momentum transfer to the intrinsic coordinates can be
neglected and the equation above can be written as%
\begin{equation}
T_{\mathrm{(DWBA)}}=\left\langle \chi^{\left( -\right) }\left( \mathbf{R}%
\right) \phi_{c}\left( \mathbf{r}\right) \left\vert U_{int}(\mathbf{r,R}%
)\right\vert \chi^{\left( +\right) }\left( \mathbf{R}\right) \phi_{i}\left(
\mathbf{r}\right) \right\rangle \ ,   \label{TDWBA}
\end{equation}
which has the known form of the DWBA T-matrix. The scattering phase space
now only depends on the center of mass momentum transfer $\mathbf{Q}$. When
the center of mass scattering waves are represented by eikonal
wavefunctions, one has%
\begin{equation}
\chi^{\left( -\right) \ast}\left( \mathbf{R}\right) \chi^{\left( +\right)
}\left( \mathbf{R}\right) \simeq S\left( b\right) \exp\left[ i\mathbf{Q}\cdot%
\mathbf{R}\right] \ .   \label{chieik}
\end{equation}
This shows that the PWBA and the DWBA are perturbative expansions of the
diffraction dissociation formula \ref{dif1}.

In DWBA (or in the eikonal approximation, eq. \ref{chieik}), $b$ does not
have the classical meaning of an impact parameter. To obtain the
semiclassical limit one goes one step further. By using eq. \ref{TDWBA0} and
assuming that $R$ depends on time so that $R=\left( \mathbf{b},Z=vt\right) $%
, the semiclassical scattering amplitude is given by $A_{\mathrm{(semiclass)}%
}^{\left( i\rightarrow f\right) }\left( b\right) =\int d^{2}b\ a_{\mathrm{%
(semiclass)}}^{\left( i\rightarrow f\right) }\left( b\right) $ exp$\left( i%
\mathbf{Q}\cdot\mathbf{b}\right) $, where%
\begin{equation}
a_{\mathrm{(semiclass)}}^{\left( i\rightarrow f\right) }\left( b\right) =%
\frac{1}{i\hbar}\ S\left( b\right) \int dtd^{3}r\ \exp\left( i\omega _{if}\
t\right) \phi_{f}^{\ast}\left( \mathbf{r}\right) U_{int}(\mathbf{r,}%
t)\phi_{i}\left( \mathbf{r}\right) \ ,   \label{semi1}
\end{equation}
where $Q_{z}Z=\omega_{if}\ t$  was used.

The semiclassical probability for the transition $\left(
i\rightarrow f\right) $\ is obtained from the above equations after
integration over \textbf{Q}. One gets
$P_{\mathrm{(semiclass)}}^{\left( i\rightarrow f\right) }\left(
b\right) =\left\vert a_{\mathrm{(semiclass)}}^{\left( i\rightarrow
f\right) }\left( b\right) \right\vert ^{2}$, with $b$ having now the
explicit meaning of an impact parameter. Thus, $a_{\mathrm{(semiclass)}%
}^{\left( i\rightarrow f\right) }\left( b\right) $ is the
semiclassical excitation amplitude. Equation \ref{semi1} is
well-known (for example in Coulomb excitation at low energies, where
$U_{int}=U_C$) except that the factor $S\left( b\right) $ is usually
set to one. In high energy collisions it is crucial to keep this
factor, as it accounts for refraction and absorption at small impact
parameters: \ $\left\vert S\left( b\right) \right\vert
^{2}=\exp\left[ 2\chi^{\mathrm{(imag)}}\right] $, where $\chi
^{\mathrm{(imag)}}$ is calculated with the imaginary part of the
optical potential. The derivation of the DWBA and semiclassical
limits of eikonal methods can be easily extended to higher-orders in
the perturbation $V$. The eikonal method includes all terms of the
perturbation series in the sudden-collision limit.

The developments presented in this section show that the DWBA
calculations of nuclear excitation, and the higher order terms, are
implicitly included in the eikonal models. However, there is a
subtle link to the optical potential, which makes the theory Lorentz
covariant. Usually, the optical limit of the Glauber-eikonal series
is used. In this model, no explicit reference to a nuclear potential
is done; only the nucleon-nucleon cross sections and nuclear
densities are used as input (see, e.g. ref. \cite{BD04}). As shown
above, this is also not a guarantee of Lorentz covariance. Even
worse is the fact that often DWBA (and higher-order, e.g. CDCC
\cite{Ra74,SYK86,NT98}) calculations are used without consideration
of relativistic effects. In the next section I give an example of a
method (relativistic CDCC) which incorporates relativistic
corrections in a continuum-discretized basis \cite{CB05}.

\section{Relativistic continuum discretized coupled-channels}
Let us consider the Klein-Gordon (KG) equation with a potential
$V_{0}$ which transforms as the time-like component of a four-vector
\cite{AC79} (here I use the notation $\hbar=c=1$). For a system with
total energy $E$ (including the rest mass $M$), the KG equation can
be cast into the form of a Schr\"{o}dinger equation (with
$\hbar=c=1$),
$\left(  \nabla^{2}+k^{2}-U\right)  \Psi=0,$ where $k^{2}=\left(  E^{2}%
-M^{2}\right)  $ and $U=V_{0}(2E-V_{0})$. When $V_{0}\ll M$, and
$E\simeq M$, one gets $U=2MV_{0}$, as in the non-relativistic case.
The condition $V_{0}\ll M$ is met in peripheral collisions between
nuclei at all collision energies. Thus, one can always write
$U=2EV_{0}$. A further simplification is to assume that the center
of mass motion of the incoming projectile and outgoing fragments is
only weakly modulated by the potential $V_{0}$. To get the dynamical
equations, one discretizes the wavefunction in terms of the
longitudinal center-of-mass momentum $k_{z}$, using the ansatz%
\begin{equation}
\Psi=\sum_{\alpha}\mathcal{S}_{\alpha}\left(  z,\mathbf{b}\right)
\ \exp\left(  ik_{\alpha}z\right)  \ \phi_{k_{\alpha}}\left(
\mathbf{\mbox{\boldmath$\xi$}}\right)  \ . \label{eq1}%
\end{equation}
In this equation, $\left(  z,\mathbf{b}\right)  $ is the
projectile's center-of-mass coordinate, with \textbf{b} equal to the
transverse coordinate. $\ \phi\left(
\mathbf{\mbox{\boldmath$\xi$}}\right) $ is the projectile intrinsic
wavefunction and $\left( k,\mathbf{K}\right) $ is the projectile's
center-of mass momentum with longitudinal momentum $k$\ and
transverse momentum $\mathbf{K}$\textbf{.} There are hidden,
uncomfortable, assumptions in eq. \ref{eq1}. The separation between
the center of mass and intrinsic coordinates is not permissible
under strict relativistic treatments. For high energy collisions we
can at best justify eq. \ref{eq1} for the scattering of light
projectiles on heavy targets. Eq. \ref{eq1} is only reasonable if
the projectile and target closely maintain their integrity during
the collision, as in the case of very peripheral collisions.

Neglecting the internal structure means $\phi_{k_{\alpha}}\left(
\mathbf{\mbox{\boldmath$\xi$}}\right)  =1$ and the sum in eq.
\ref{eq1} reduces to a single term with $\alpha=0$, the projectile
remaining in its ground-state. It is straightforward to show that
inserting eq. \ref{eq1} in the KG equation $\left(
\nabla^{2}+k^{2}-2EV_{0}\right)  \Psi=0$, and neglecting
$\nabla^{2}\mathcal{S}_{0}\left(  z,\mathbf{b}\right)  $ relative to
$ik\partial_{Z}\mathcal{S}_{0}\left(  z,\mathbf{b}\right)  $, one
gets
$ik\partial_{Z}\mathcal{S}_{0}\left(  z,\mathbf{b}\right)  =EV_{0}%
\mathcal{S}_{0}\left(  z,\mathbf{b}\right)  $, which leads to the
center of mass scattering solution $\mathcal{S}_{0}\left(
z,\mathbf{b}\right) =\exp\left[
-iv^{-1}\int_{-\infty}^{z}dz^{\prime}\ V_{0}\left(  z^{\prime
},\mathbf{b}\right)  \right]  ,$ with $v=k/E$. Using this result in
the Lippmann-Schwinger equation, one gets the familiar result for
the eikonal elastic scattering amplitude, i.e. $f_{0}=-i\left( k/
2\pi\right) \int d\mathbf{b}\ \exp\left(  i\mathbf{Q\cdot b}\right)
\ \left\{  \exp\left[ i\chi(\mathbf{b})\right]  -1\right\}  , $
where the eikonal phase is given by
$\exp[i\chi(\mathbf{b})]=\mathcal{S}_{0}\left(
\infty,\mathbf{b}\right) $, and
$\mathbf{Q}=\mathbf{K}^{\prime}-\mathbf{K}$ is the transverse
momentum transfer. \ Therefore, the elastic scattering amplitude in
the eikonal approximation has the same form as that derived from the
Schr\"odinger equation in the non-relativistic case.

For inelastic collisions we insert eq. \ref{eq1} in the KG equation
and use the orthogonality of the intrinsic wavefunctions
$\phi_{k_{\alpha}}\left( \mathbf{\mbox{\boldmath$\xi$}}\right)  $.
This leads to a set of coupled-channels equations for
$\mathcal{S}_{\alpha}$:%
\begin{equation}
\left(  \nabla^{2}+k^{2}\right)
\mathcal{S}_{\alpha}\mathrm{e}^{ik_{\alpha}
  z}=\sum_{\alpha }\left\langle \alpha\left\vert U\right\vert
\alpha^{\prime}\right\rangle \ \mathcal{S}_{\alpha^{\prime}}\
\mathrm{e}^{i
k_{\alpha^{\prime}}  z}, \label{eq2}%
\end{equation}
\bigskip with the notation $\left\vert \alpha\right\rangle =\left\vert
\phi_{k_{\alpha}}\right\rangle $. Neglecting terms of the form $\nabla
^{2}\mathcal{S}_{\alpha}\left(  z,\mathbf{b}\right)  $ relative to
$ik\partial_{Z}\mathcal{S}_{\alpha}\left(  z,\mathbf{b}\right)  $, eq.
\ref{eq2} reduces to%
\begin{equation}
iv\frac{\partial\mathcal{S}_{\alpha}\left(  z,\mathbf{b}\right)  }{\partial
z}=\sum_{\alpha^{\prime}}\left\langle \alpha\left\vert V_{0}\right\vert
\alpha^{\prime}\right\rangle \ \mathcal{S}_{\alpha^{\prime}}\left(
z,\mathbf{b}\right)  \ \mathrm{e}^{i\left(  k_{\alpha^{\prime}}-k_{\alpha%
}\right)  z}. \label{eq3}%
\end{equation}
The scattering amplitude for the transition $0\rightarrow\alpha$ is given by%
\begin{equation}
f_{\alpha}\left(  \mathbf{Q}\right)  =-\frac{ik}{2\pi}\int d\mathbf{b}%
\ \ \exp\left(  i\mathbf{Q\cdot b}\right)  \ \left[  S_{\alpha}\left(
\mathbf{b}\right)  -\delta_{\alpha,0}\right]  , \label{eq4}%
\end{equation}
with $S_{\alpha}\left(  \mathbf{b}\right)
=\mathcal{S}_{\alpha}\left( z=\infty,\mathbf{b}\right)  $. The set
of equations \ref{eq3} and \ref{eq4} are the relativistic-CDCC
equations (RCDCC).

The RCDCC equations have been used \cite{CB05} to study the
dissociation of $^{8}$B projectiles at high energies. The energies
transferred to the projectile are small, so that the wavefunctions
can be treated non-relativistically in the projectile frame of
reference. In this frame the wavefunctions are described in
spherical coordinates, i.e. $\left\vert \alpha\right\rangle
=\left\vert jlJM\right\rangle $, where $j$, $l$, $J$ and $M$ denote
the angular momentum numbers characterizing the projectile state.
Eq. \ref{eq3} is Lorentz invariant if the potential $V_0$ transforms
as the time-like component of a four-vector. The matrix element
$\left\langle \alpha\left\vert V_0\right\vert
\alpha^{\prime}\right\rangle $ is also Lorentz invariant, and one
can therefore calculate them in the projectile frame.

The longitudinal wavenumber $k_{\alpha}\simeq(E^{2}-M^{2})^{1/2}$\
also defines how much energy is gone into projectile excitation,
since for small energy and momentum transfers
$k_{\alpha}^{\prime}-k_{\alpha}=\left(
E_{\alpha}^{\prime}-E_{\alpha}\right) /v$. In this limit, eqs.
\ref{eq3} and \ref{eq4} reduce to semiclassical coupled-channels
equations, if one uses $z=vt$ for a projectile moving along a
straight-line classical trajectory, and changing to the notation
$\mathcal{S}_{\alpha}\left(  z,b\right) =a_{\alpha }(t,b)$, where
$a_{\alpha}(t,b)$ is the time-dependent excitation amplitude for a
collision wit impact parameter $b$ (see eqs. 41 and 76 of ref.
\cite{BCG03}). The full version of eq. \ref{eq4} was used in ref.
\cite{CB05}, with relativistic corrections in both the Coulomb and
nuclear potentials.

If the state $\left\vert \alpha\right\rangle $\ is in the continuum (positive
proton+$^{7}$Be energy) the wavefunction is discretized according to
$\left\vert \alpha;E_{\alpha}\right\rangle =\int dE_{\alpha}^{\prime}%
\ \Gamma(E_{\alpha}^{\prime})\ \left\vert \alpha;E_{\alpha}^{\prime
}\right\rangle $, where the functions $\Gamma(E_{\alpha})$ are
assumed to be strongly peaked around the energy $E_{\alpha}$ with
width $\Delta E$. For convenience the histogram set (eq. 3.6 of ref.
\cite{BC92}) is chosen. The inelastic cross section is obtained by
solving the RCDCC equations and using $d\sigma/d\Omega dE_{\alpha
}=\left\vert f_{\alpha}\left( \mathbf{Q}\right)  \right\vert ^{2}\
\Gamma ^{2}(E_{\alpha})$.

\begin{figure}[ptb]
\includegraphics[height=.3\textheight]{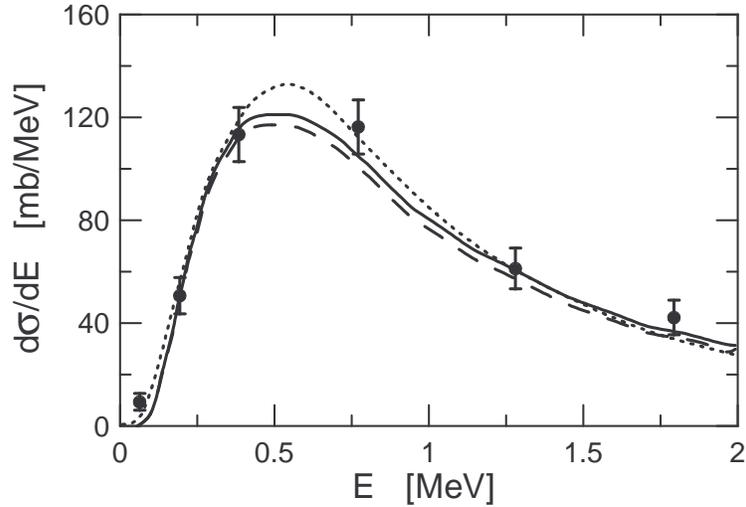}
\caption{ Cross sections for the dissociation reaction $^{8}$B+Pb
$\rightarrow$ p+$^{7}$Be+Pb at 83 MeV/nucleon and for
$\theta_{8}<1.8^{0}$. Data are from ref. \cite{Dav01}. The dotted
curve is the first-order perturbation result. The solid curve is the
RCDCC calculation. The dashed curve is obtained with the replacement
of $\gamma$ by unity in the nuclear and
Coulomb potentials.}%
\label{f2}%
\end{figure}

Figure \ref{f2} shows the relative energy spectrum between the proton and the $^{7}%
$Be after the breakup of $^{8}$B on lead targets at 83 MeV/nucleon.
The data are from ref. \cite{Dav01}. In this case, the calculation
was restricted to $b>30$ fm. The dotted curve is the first-order
perturbation calculation, the solid curve is the RCDCC calculation,
and the dashed curve is obtained with the replacement of $\gamma$ by
unity in the nuclear and Coulomb potentials. The difference between
the solid and the dashed-curve is of the order of 4-9\%.

\section{Conclusions}

The consequence of neglecting relativity in nuclear reactions at
intermediate energies is not easy to access. The inclusion of
relativity introduces non-trivial effects in semiclassical, DWBA,
eikonal, and continuum discretized coupled-channels calculations.
Nuclear collisions are up to now the most used probe of the internal
structure of rare nuclear isotopes. To my knowledge, most
experiments have been analyzed using non-relativistic theoretical
methods. It might be necessary to review the results of some of
these data, using a proper treatment of the relativistic corrections
in the theoretical calculations used in the experimental analysis.
Other improvements of the formalisms presented here and elsewhere
needs to be assessed. The relativistic effects in the nuclear
interaction has also to be studied in more depth.

Special relativity \cite{Ei1905}, one of the most precious theories
of Eintein's legacy, still remains a source of intriguing effects,
not always easy to tackle. Nuclear physics is full of such examples.

\begin{theacknowledgments}
I wish to thank Gerhard Baur, Kai Hencken and Stefan Typel for
discussions on various topics in this field. This work was supported
by the U.\thinspace S.\ Department of Energy under grant No.
DE-FG02-04ER41338.
\end{theacknowledgments}

\end{document}